\documentclass[letters,fleqn,usenatbib]{mnras}

\usepackage{newtxtext,newtxmath}

\usepackage[T1]{fontenc}
\usepackage{ae,aecompl}

\usepackage{graphicx}	
\usepackage{amsmath}	
\usepackage{amssymb}	
\usepackage{bm}
\usepackage[colorinlistoftodos]{todonotes}
\usepackage{dsfont}

\DeclareMathOperator{\n}{\text{n}}
\DeclareMathOperator{\p}{\text{p}}

\DeclareMathOperator{\cru}{\, \text{cr}}
\DeclareMathOperator{\cop}{\text{pin}} 
\DeclareMathOperator{\conp}{\text{f}}

\title[Vortex pinning and glitch rises]{Vortex pinning in the superfluid core of neutron stars and the rise of pulsar glitches}

\author[A. Sourie and N.~Chamel]{
Aur\'elien Sourie,$^{1,2}$\thanks{E-mail: asourie@ulb.ac.be}
and Nicolas Chamel$^{1}$\thanks{E-mail: nchamel@ulb.ac.be}
\\
$^{1}$Institut d'Astronomie et d'Astrophysique, CP-226, Universit\'e Libre de Bruxelles, 1050 Brussels,
Belgium\\
$^{2}$Laboratoire Univers et Th\'eories, Observatoire de Paris, PSL Research University, CNRS, Universit\'e Paris Diderot, \\  Sorbonne Paris Cit\'e,  5 place Jules Janssen, 92195 Meudon, France\\
}

\date{Accepted XXX. Received YYY; in original form ZZZ}

\pubyear{2020}

\begin{document}
\label{firstpage}
\pagerange{\pageref{firstpage}--\pageref{lastpage}}
\maketitle

\begin{abstract}
Timing of the Crab and Vela pulsars have recently revealed very peculiar evolutions of their spin frequency during the early stage of a glitch. We show that these differences can be interpreted from the interactions between neutron superfluid vortices and proton fluxoids in the core of these neutron stars. In particular, pinning of individual vortices to fluxoids is found to have a dramatic impact on the mutual friction between the neutron superfluid and the rest of the star. The number of fluxoids attached to vortices turns out to be  a key parameter governing the global dynamics of the star. These results may have implications for the interpretation of other astrophysical phenomena such as pulsar-free precession or the r-mode instability. 
\end{abstract}

\begin{keywords}
stars:neutron -- pulsars: individual (PSR B0833$-$45, PSR B0531+21)
\end{keywords}


\section{Introduction}
\label{sec:intro}

Pulsars are neutron stars (NSs) spinning very rapidly with extremely stable periods. With relative delays as small as $10^{-21}$, some pulsars outperform the most accurate terrestrial clocks~\citep{milner2019}. Nevertheless, irregularities have been detected in long-term pulsar timing observations. In particular, some pulsars have been found to suddenly spin up. Such `glitches' in their rotational frequency $\Omega$, ranging from $\Delta\Omega\slash \Omega\sim 10^{-9}$ to $\sim 10^{-5}$, are sometimes accompanied by an abrupt change of the spin-down rate from $\vert\Delta \dot\Omega\slash\dot\Omega\vert\sim 10^{-6}$ up to $\sim 10^{-2}$~\citep{manchester2018}. At the time of this writing, 554 glitches have been detected in 191 pulsars\footnote{\url{http://www.jb.man.ac.uk/pulsar/glitches/gTable.html}}~\citep{espinoza2011}. The very long post-glitch relaxation, lasting from days to years, reveals the presence of superfluid components in NSs~\citep{chamel2017}. Glitches themselves are thought to be the manifestations of superfluidity~\citep{haskell2015models}. These events are commonly interpreted as sudden transfers of angular momentum from a more rapidly rotating neutron superfluid to the rest of star due to the catastrophic unpinning of quantised vortices. However, large uncertainties remain concerning the dynamics of these vortices. In particular, protons in the outer core of a NS are generally thought to form a type-II superconductor such that the magnetic flux penetrates through fluxoids, each carrying a magnetic flux quantum $\phi_0=hc/(2e)\simeq 2\times 10^{-7} \text{ G~cm}^{2}$ where $h$ is Planck's constant, $c$ the speed of light and $e$ the proton electric charge. The mean surface density of fluxoids, $\mathcal{N}_{\p} \simeq 5\times 10^{18}\, B_{12}$~cm$^{-2}$ where $B_{12} = B / 10^{12}\text{\ G}$ is the stellar internal magnetic field, is huge compared to that of vortices, $\mathcal{N}_{\n} \simeq 6\times 10^5 / P_{10}$~cm$^{-2}$ where $P_{10}= P / 10\text{ ms}$ is the observed rotation period. Vortices may pin to fluxoids, and this may affect significantly the dynamical evolution of the star~\citep{alpar2017}. Nevertheless, the role of the core superfluid on the glitch rise remains to be investigated. 

So far, the most detailed information come from the large glitches recently detected in the Vela~\citep{palfreyman18alteration,asthon19rotational} and Crab pulsars~\citep{shaw2018largest}, revealing very different behaviours. The analysis of the Vela glitch observed in Dec. 2016 suggests the presence of an overshoot of amplitude\footnote{The amplitude $\Delta f_{\text{over}}$ given here corresponds to the magnitude of the exponentially-decaying term plus that of the final frequency jump, respectively denoted by $\Delta f_\text{d}$ and $\Delta f$ in~\citet{asthon19rotational}.} $\Delta f_{\text{over}} \sim 19-38$~$\mu$Hz, significantly larger than the amplitude of the pulsar frequency jump at the end of the rise stage, $\Delta f\simeq 16$~$\mu$Hz. While the timescale $\tau_\text{r}$ associated with the glitch rise is found to be shorter than $\sim12$~s, a longer timescale has been deduced for the subsequent decrease, $\tau_\text{d}\sim 41-125$~s. These two timescales are  compatible with observations of previous Vela glitches~\citep{dodson02high,dodson2007two}.  Furthermore, some evidence for the existence of a precursor (in the form of a rapid slow-down preceding the glitch) may have been found in the 2016 Vela glitch. On the other hand, a delayed spin-up, consisting of a first unresolved frequency jump $\Delta f_\text{short}$ over a short timescale $\tau_\text{short}$ followed by a resolved spin-up with an amplitude $\Delta f_\text{long}$ over a longer timescale $\tau_\text{long}$, have been detected in the 1989, 1996 and 2017 Crab glitches~\citep{lyne92spin, wong2001observations, shaw2018largest}. The analysis of the 2017 Crab glitch has led to $\Delta f_\text{short}\simeq 14$~$\mu$Hz, $\Delta f_\text{long}\simeq 1.1$~$\mu$Hz, $\tau_\text{short} \leq 0.45$~d and $\tau_\text{long}\simeq 1.7$~d, corresponding to a total amplitude $\Delta f = \Delta f_\text{short} + \Delta f_\text{long} \simeq 15$~$\mu$Hz, and a total rise time $\tau_\text{r} \sim \tau_\text{short} + \tau_\text{long}\sim 2$~d. Similar timescales have been deduced from the analyses of the 1989 and 1996 glitches, but amplitudes $\sim10$ times smaller were observed. Finally, let us stress that the spin-up stage has not been resolved for smaller Crab glitches.

The glitch rise is thought to be governed by mutual-friction forces between the superfluid and the rest of the star, arising from the dissipative forces acting on individual vortices~\citep{haskell2015models,sourie2017global,graber2018,haskell2018}. Recently, \cite{haskell2018} have suggested that the different spin-up evolutions observed in the Vela and Crab pulsars could be explained by the different stellar regions (core vs crust) where the glitch is driven. In this Letter, we explore the impact of vortex pinning only in the outer core  of NSs on the glitch rise.

\section{Smooth-averaged hydrodynamic description}

\subsection{Forces on a single vortex} 
\label{sec:single_vortex}

Let us consider a single neutron  vortex pinned to $N_{\p}$ proton fluxoids and moving with velocity $v_{\mathrm{L}}^i$ ($i=1,2,3$ denoting spatial indices). The vortex is assumed to be evolving in a mixture of superconducting protons, (degenerate) electrons and superfluid neutrons at zero temperature. Although the arrangement of fluxoids in the core of a NS may be quite complicated, depending not only on the cooling and magneto-rotational evolution of the star~\citep{srinivasan1990novel,ruderman1998neutron,jahan-miri2000flux} but also on the nature of the phase transition~\citep{haber2017}, we suppose for simplicity that the $N_{\p}$ pinned fluxoids are aligned with the vortex~\citep{ding1993magnetic,ruderman1998neutron}. This assumption is actually not completely unrealistic, at least at small enough scales~\citep{drummond2017}.  Note that the pinned fluxoids  are not necessarily superimposed on the vortex. Vortex-fluxoid clusters may actually form naturally~\citep{sedrakian1995superfluid}. Here, $N_{\p}$ is an unknown parameter that could potentially be as large as $N_{\p}^{\text{max}} \sim \mathcal{N}_{\p}/\mathcal{N}_{\n} \simeq 10^{13}  \, B_{12} \, P_{10}$. We further assume that the vortex is straight, infinitely rigid and we ignore the effects of gravity giving rise to a buoyancy force (see \cite{dommes2017vortex} for a recent discussion).

We determine the force felt by a single vortex moving in an asymptotically uniform superfluid mixture following an approach originally developed by \cite{carter2002relativistic} in the relativistic framework, and later adapted to the Newtonian context by \cite{carter2005covariant}. Making use of the results obtained by \cite{gusakov2019} for electrons, the force acting on the vortex can be decomposed into three parts: $\mathcal{F}^{i} = \mathcal{F}^{i}_{\mathrm{M} \n} +  \mathcal{F}^{i}_{\mathrm{M} \p} + \mathcal{F}^{i}_{\mathrm{d}}\, $, as shown in an accompanying paper~\citep{sourie20}. 
The neutron Magnus force arising from the relative flow of superfluid neutrons with velocity $v_{\n}^i$ - $v_{\mathrm{L}}^i$ is given by 
\begin{equation}
\label{Magnus_n}
\mathcal{F}^{i}_{\mathrm{M} \n} = - \rho_{\n}\, \varepsilon^{ijk}\,  \kappa \,  \hat{\kappa}_j \, \left( v_{\n k} - v_{\mathrm{L}\,  k} \right)\, , 
\end{equation}
where $\rho_{\n}$ is the neutron mass density, $\kappa = h/(2\, m)$ is the quantum of circulation ($m$ denoting the neutron rest mass, taken to be equal to that of protons) and $\hat{\kappa}^i$ is a unit vector oriented along the vortex.  Likewise, the   flow of protons  with velocity $v_{\p}^i-v_{\mathrm{L}}^i$ relative to pinned fluxoids leads to a Magnus type force
\begin{equation}
\label{force_Magnus_proton}
\mathcal{F}^{i}_{\mathrm{M} \p} = - \rho_{\p}\, N_{\p}\,  \varepsilon^{ijk}\, \kappa \, \hat{\kappa}_j\, \left( v_{\p k} - v_{\mathrm{L}\,  k} \right)\, ,
\end{equation}
where $\rho_{\p}$ is the proton mass density. The scattering of electrons off the magnetic field carried by fluxoids, and to a lesser extent that induced by entrained protons around the vortex~\citep{alpar1984rapid}, leads to the drag force \begin{equation}
\label{dragNewt}
\mathcal{F}^{i}_{\mathrm{d}} = - \rho_{\n}\, \kappa \, \xi \,  \left( v_{\mathrm{L}}^i - v_{\p}^i \right)\, ,
\end{equation}
where $\xi >0$ is the so-called drag-to-lift ratio.

\subsection{Global averaging over many vortices}
\label{sec:hydro}

On length scales much larger than the intervortex separation, the electrically charged particles inside NSs are strongly coupled and essentially co-rotate with the crust and the magnetosphere~\citep{glampedakis2011magnetohydro}. The outer core of a NS can therefore be reasonably well described by means of a two-fluid model, involving (i) a neutron superfluid moving with velocity $v_{\n}^i$ and (ii) a (viscous) charge-neutral fluid made of  protons and electrons (simply labelled by '$\p$' in the following), moving with velocity $v_{\p}^i$. The two fluids are mutually coupled by friction forces induced by the drag force~\eqref{dragNewt}. The smooth-averaged force per unit volume exerted by the vortices on the superfluid (ignoring interactions between vortices) is given by 
\begin{equation}
\label{def_mf_magnus}
f_{\text{mf}}^i= - \mathcal{N}_{\n} \, \mathcal{F}^{i}_{\mathrm{M}\n }\, .
\end{equation}

Solving the force balance equation of a single vortex (neglecting its mass) $\mathcal{F}^{i}_{\mathrm{d}} + \mathcal{F}^{i}_{\mathrm{M} \n} + \mathcal{F}^{i}_{\mathrm{M} \p} = 0$ for the vortex velocity $v_{\mathrm{L}}^i$ following standard procedure~\citep{hall1956vinen} and substituting into~\eqref{def_mf_magnus} yield 
\begin{equation}
\label{MFNewt2}
f_{\text{mf}}^i= - \mathcal{N}_{\n}\, \rho_{\n} \, \kappa \, \left(\mathcal{B}' \,\varepsilon^{ijk}\hat{\kappa}_j w^{\p\!\n}_k + \mathcal{B}  \,\varepsilon^{ijk}\hat{\kappa}_j\varepsilon_{klm}\hat{\kappa}^l w_{\p\!\n}^m  \right)\, ,
\end{equation}
where $w_{\p\!\n}^i = v_{\p}^i - v_{\n}^i$, 
\begin{equation}
\label{beta}
\mathcal{B}=  \dfrac{\xi}{\left(1+X\right)^2 + \xi^{2}} \, , \   \mathcal{B}' = 1-\dfrac{1+X}{\left(1+X\right)^2 + \xi^{2}}~, \  X= \dfrac{x_{\p}}{1-x_{\p}}N_{\p}\, ,
\end{equation}
$x_{\p}=\rho_{\p}/(\rho_{\p}+\rho_{\n})$ 
denoting the proton fraction. While expression~\eqref{MFNewt2} is formally similar to that obtained in the absence of pinning~\citep{mendell1991superfluid}, pinning is found to affect the actual values of the mutual-friction coefficients~\eqref{beta} by (i) modifying the drag-to-lift ratio $\xi$, and (ii) inducing an extra dependence on $X$ due to the proton Magnus force~\eqref{force_Magnus_proton}. The drag force remains poorly known. If the $N_{\p}$ fluxoids and the vortex are superimposed, $\xi\propto N_{\p}^2$~\citep{ding1993magnetic} while $\xi\propto N_{\p}$ according to the vortex-cluster model of~\citet{sedrakian1995superfluid}. Given the current lack of knowledge, we adopt the following parametrization\footnote{The factor of $1/\varepsilon_{\p}^2$ is needed so as to recover known results for a single fluxoid~\citep{gusakov2019}.}:
\begin{equation}
\label{eq:xi_ancrage}
    \xi =\xi_0 \times \left(\varepsilon_{\p}\right)^{-2} \times  \left(N_{\p}\right)^{\alpha}  \ \ \mbox{ if $N_{\p} > 0$}\, , 
\end{equation}
where $\xi_0$ is the drag-to-lift ratio in the absence of pinning~\citep{alpar1984rapid, mendell1991superfluid, andersson2006mutual}
\begin{equation}
\label{eq:xi_coeur_nonpinned}
       \xi_0 = 4\times 10^{-4}\dfrac{ \varepsilon_{\p}^2}{\left(1- \varepsilon_{\p}\right)^{1/2}} \left(\dfrac{x_{\p}}{0.05}\right)^{7/6}\dfrac{1}{1-x_{\p}}   {\rho_{14}}^{1/6}\, ,
\end{equation}
$\varepsilon_{\p}$ denoting the proton entrainment parameter, 
and $\rho_{14}=\rho/(10^{14}~{\rm g~cm}^{-3})$ the mass density. 
\begin{figure}
 \center 
   \includegraphics[width = \columnwidth]{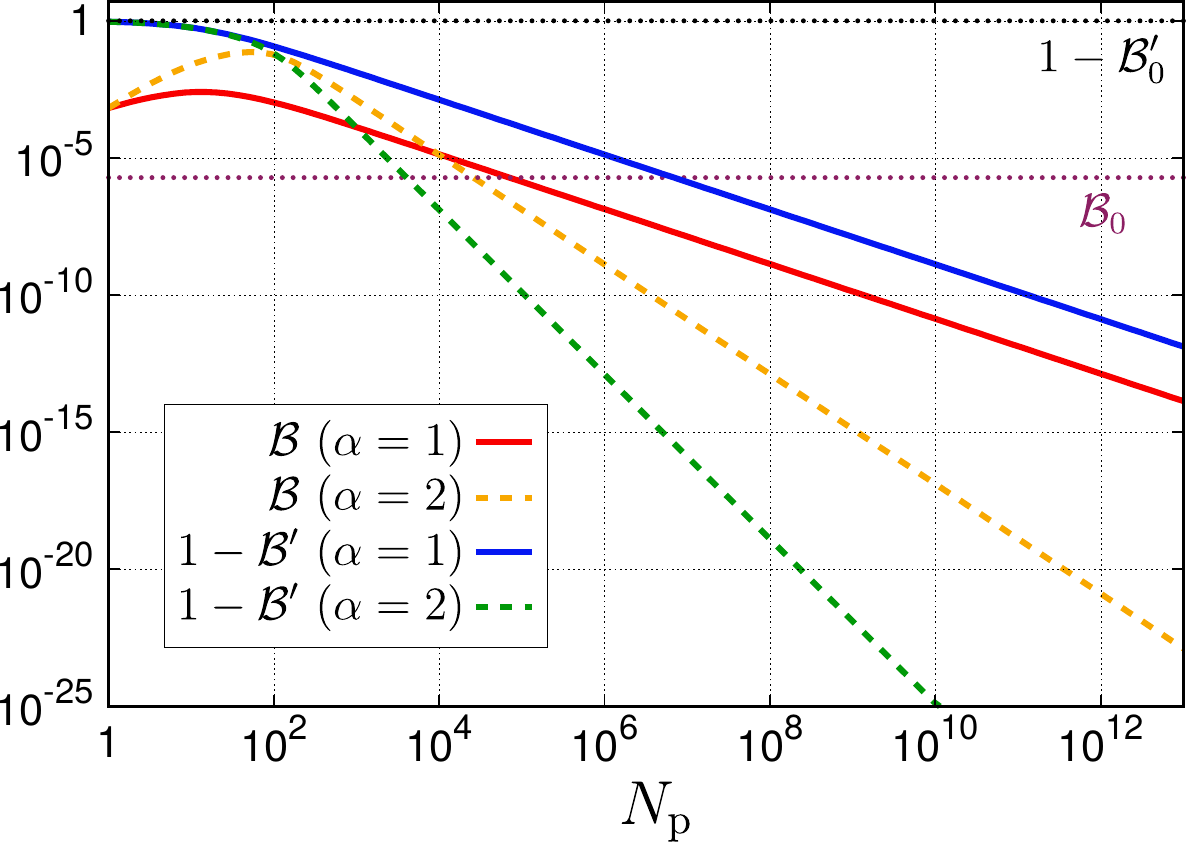}  
  \caption{(Colors online) Mutual-friction coefficients $\mathcal{B}$ and $1-\mathcal{B}'$ in the outer core of NSs as functions of the number $N_{\p}$ of pinned fluxoids for $\alpha = 1$ (solid lines) and $\alpha = 2$ (dashed lines). Corresponding values for $N_{\p} = 0$ are indicated as horizontal lines. 
   }
 \label{fig:beta}
\end{figure}

The mutual-friction coefficients are plotted in Fig.~\ref{fig:beta}. The following typical values for the other parameters were adopted: $\varepsilon_{\p} = 0.05$, $x_{\p} = 0.07$ and $\rho_{14} = 2.7$. The mutual-friction coefficients $\mathcal{B}_0\equiv \mathcal{B}\left(N_{\p}=0\right) $ and $\mathcal{B}'_0\equiv \mathcal{B}'\left(N_{\p}=0\right)$ in the absence of pinning are displayed by horizontal lines in Fig.~\ref{fig:beta}. For both values of $\alpha$, $\mathcal{B}\gg \mathcal{B}_0$ for small enough  values of $N_{\p}$, while the opposite behaviour is observed at higher $N_{\p}$. Moreover, $\mathcal{B}'\simeq \mathcal{B}'_0 \simeq 0$ for $N_{\p}\gtrsim 0$, while $\mathcal{B}' \simeq 1$ at higher $N_{\p}$. Pinning may thus have a dramatic impact on the mutual-friction force and on the superfluid dynamics of NSs depending on $N_{\p}$. Similar conclusions can be drawn for any real value of $\alpha$. 

\section{Astrophysical implications for pulsar glitches}
\label{astro}

\subsection{Minimal model}

To investigate the impact of core vortex pinning on the glitch dynamics, we consider a `minimal' model in which the NS is simply described in terms of three dynamically distinct components: (i) a `pinned' neutron superfluid in the outer core where the magnetic field is predominantly toroidal and pinning to fluxoids is expected to be the most effective~\citep{haskell2013,gugercinoglu2014}, (ii) a `non-pinned' neutron superfluid in the inner core, and (iii) the rest of the star. In view of the strong entrainment in the crust~\citep{chamel2012neutron}, we assume for simplicity that only the core neutron superfluid participates to the glitch. The third component, simply referred to as `proton' in the following, thus consists of all charged particles (protons, leptons, nuclei in the crust) and the crustal neutron superfluid. All three components are rigidly rotating around a common axis, $z$ say, at the angular velocity $\Omega_{\n}^{\cop}$, $\Omega_{\n}^{\conp}$  and $\Omega_{\p}$, respectively. The corresponding moments of inertia are denoted by $I_{\n}^{\cop}$, $I_{\n}^{\conp}$ and $I_{\p}$, and satisfy $I_{\n}^{\cop} + I_{\n}^{\conp}+ I_{\p} = I$, where $I$ is the total moment of inertia of the star. Due to magnetic couplings, the proton component  essentially rotates at the observed pulsar angular velocity $\Omega$.

We further assume that the pinned and non-pinned core superfluids are dynamically coupled to the proton fluid through mutual friction only. Although such a simple picture is a priori inadequate to describe the long-term post-glitch relaxation (for which additional processes such as vortex creep occur), our model can nevertheless be safely applied to the short spin-up stage. For simplicity, the mutual-friction coefficients associated with the pinned and non-pinned core superfluids, respectively denoted by $\mathcal{B}_{\cop}$ and $\mathcal{B}_{\conp}$, are supposed to be uniform and time independent. In other words, each vortex in the pinned region remains anchored to the same number $N_{\p}$ of fluxoids during the glitch rise. Mutual friction between the proton fluid and the core superfluid ${\textrm{\scriptsize $X$}}$ is accounted for through the torque $\Gamma^i_{_X} = \int_{_X} \varepsilon^{ijk} x_j  f_{{_X}\, k} \, \text{d}^3V$, 
where $x^i = r \, \delta^i_r$ in spherical coordinates, $f_{_X}^{k}$ is the relevant mutual-friction force~\eqref{MFNewt2} and the integral is taken over the region ${\textrm{\scriptsize $X$}}$ under consideration. Neglecting entrainment effects between the fluids, and assuming circular motion, the $z-$component of the torque simply reads
$\Gamma_{_X}^z  = 2\, \mathcal{B}_{_X} \, I_{\n}^{_X} \, \Omega_{\n}^{_X} \left(\Omega_{\p} - \Omega_{\n}^{_X}\right)$. The dynamics of the glitch rise is thus governed by the following equations:
\begin{align}
&\dot{\Omega}_{\p} = - \dfrac{I_{\n}^{\conp}}{I_{\p}} \, \dot{\Omega}_{\n}^{\conp} - \dfrac{I_{\n}^{\cop}}{I_{\p}} \,  \dot{\Omega}_{\n}^{\cop} +  \dfrac{\Gamma_\text{ext}}{I_{\p}} \, , \label{eq_transfert1}\\[0.1 cm]
  &  \dot{\Omega}_{\n}^{\conp}  = 2\, \mathcal{B}_{\conp} \, \Omega_{\n}^{\conp} \left(\Omega_{\p} - \Omega_{\n}^{\conp}\right)\, , \label{eq_transfert2}\\[0.1 cm]
   & \dot{\Omega}_{\n}^{\cop}  = 2\, \mathcal{B}_{\cop} \, \Omega_{\n}^{\cop} \left(\Omega_{\p} - \Omega_{\n}^{\cop}\right)\, , \label{eq_transfert3} 
\end{align}
where $\Gamma_\text{ext}= I \dot{\Omega}_\infty$ stands for the external torque responsible for the slow braking of the pulsar on long timescales with spin-down rate $\dot{\Omega}_\infty$.

\subsection{Initial conditions and physical ingredients}

In view of the lack of knowledge on the pre-glitch evolution, we simply assume that the proton component and the non-pinned core neutron superfluid at the beginning of the glitch ($t=0$) are rotating with a lag corresponding to the asymptotic post-glitch steady-state lag\footnote{These post-glitch steady-state lags are obtained by imposing $\dot{\Omega}_{\n}^{\conp}=\dot{\Omega}_{\n}^{\cop} = \dot{\Omega}_\infty$ in Eqs.~\eqref{eq_transfert2} and~\eqref{eq_transfert3}.}:  $\Omega_{\n}^{\conp}(0) = \Omega_0 +  |\dot{\Omega}_\infty|/ \left(2\, \mathcal{B}_{\conp} \, \Omega_0 \right)$ where  $\Omega_0=\Omega_{\p}(0)$ (see, e.g., \cite{pizzochero19}). 
On the other hand, the initial rotation rate of the pinned core neutron superfluid is supposed to be given by $\Omega_{\n}^{\cop}(0) = \Omega_0 +   |\dot{\Omega}_\infty|/\left(2\, \mathcal{B}_{\cop} \, \Omega_0\right) + \delta\Omega_0$, where $\delta\Omega_0$ denotes a small deviation to the post-glitch steady-state lag. 

To solve Eqs.~\eqref{eq_transfert1}$-$\eqref{eq_transfert3}, the  pulsar rotation rate $\Omega_0$, the long-term spin-down rate $\dot{\Omega}_\infty$, the initial lag $\delta \Omega_0$, the mutual-friction coefficients $\mathcal{B}_{\conp}$ and  $\mathcal{B}_{\cop}$, and the ratios $I_{\n}^{\conp}/I$ and  $I_{\n}^{\cop}/I$ need to be specified. In what follows, $\Omega_0$ and $\dot{\Omega}_\infty$ are directly taken from pulsar timing. The coefficient $\mathcal{B}_{\conp}$ in the non-pinned region is given by $\mathcal{B}_0$, and the corresponding drag-to-lift ratio by Eq.~\eqref{eq:xi_coeur_nonpinned}. In the pinned region, the coefficient $\mathcal{B}_{\cop}$ is given by Eq.~\eqref{beta}, with the prescription~\eqref{eq:xi_ancrage} and suitable parameters. Typical values for the underlying parameters are: $\varepsilon_{\p}^{\cop}\simeq 0.05-0.2 $, $\varepsilon_{\p}^{\conp}\simeq 0.1-0.5 $ (see, e.g., \cite{chamel2006entrainment,sourie2016numerical}),  $x_{\p}^{\cop} \simeq 0.05 - 0.1$, $x_{\p}^{\conp} \simeq 0.05 - 0.4$, $\rho_{\cop} \simeq (0.5-2) \rho_0$ and $\rho_{\conp} \simeq (2-6) \rho_0$, $\rho_0\simeq 2.7\times 10^{14}~\text{ g~cm}^{-3}$ being the nuclear saturation
density~(see, e.g., \cite{pearson2018}). The ratios $I_{\n}^{\cop}/I$ and $I_{\n}^{\conp}/I$ are computed using the relations $I_{\n}^{_X}  /I^{_X} =  1-x_{\p}^{_X}$ (assuming uniform densities in each region), where $I^{_X}$ is the total moment of inertia of region ${\textrm{\scriptsize $X$}}$, and $I^{\conp} = I-I^{\cru}-I^{\cop}$, $I^{\cru}$ denoting the crustal moment of inertia of the star. Typical values are: $I^{\cru}/I\simeq 0.01-0.05$~\citep{delsate2016giant} and $I^{\cop}/I\sim 0.05$~\citep{gugercinoglu2014}. Unlike the previous quantities, both the initial lag $ \delta \Omega_0$ and number $N_{\p}$ of pinned fluxoids are essentially unknown. As shown in the next section, the large range of possible values for $N_{\p}$ could account for the very different spin-up behaviours in the Crab and Vela pulsars. 

\subsection{Applications to the Crab and Vela pulsars}

As discussed in the Supplementary Material (SM), the set of equations~\eqref{eq_transfert1}$-$\eqref{eq_transfert3} can be solved analytically provided that the variations of the separate angular velocities are neglected with respect to those of the lags between the fluids appearing in the right-hand side of the equations\footnote{This assumption is well-justified given the very small observed glitch amplitudes.} (see also~\cite{pizzochero19}). This analytical solution also  allows for an unambiguous definition of the relevant timescales governing the dynamics of the glitch rise.  The adopted values for the different parameters are:  $\varepsilon_{\p}^{\cop}=0.05$, $x_{\p}^{\cop} = 0.07$,  $\rho_{\cop} = \rho_0$, $\varepsilon_{\p}^{\conp}=0.1$, $x_{\p}^{\conp} = 0.2$, $\rho_{\conp} = 3 \rho_0$,  $I^{\cru}/I=0.03$ and  $I^{\cop}/I=0.08$. This choice leads to $I_{\p}/I\simeq 0.21$, $I_{\n}^{\conp}/I\simeq 0.71 $ and $I_{\n}^{\cop}/I\simeq 0.08$. The initial pulsar frequency $\Omega_0/2\pi$ is fixed to $11.19$~Hz (resp. $29.64$~Hz) for the Vela (resp. Crab) pulsar~\citep{dodson2007two, shaw2018largest}. Focusing on the deviation $\Delta \Omega_{\p}(t) = \Omega_{\p}(t)-\Omega_\text{pre}(t)$ induced by the glitch event in the evolution of the pulsar rotation rate, where  $\Omega_\text{pre}(t) = \Omega_0 + \dot{\Omega}_\infty  t$ is the rotation rate extrapolated from the pre-glitch evolution, the actual value of $\dot{\Omega}_\infty$ is unimportant.

\begin{figure*}
 \center 
   \includegraphics[width = 0.99\columnwidth]{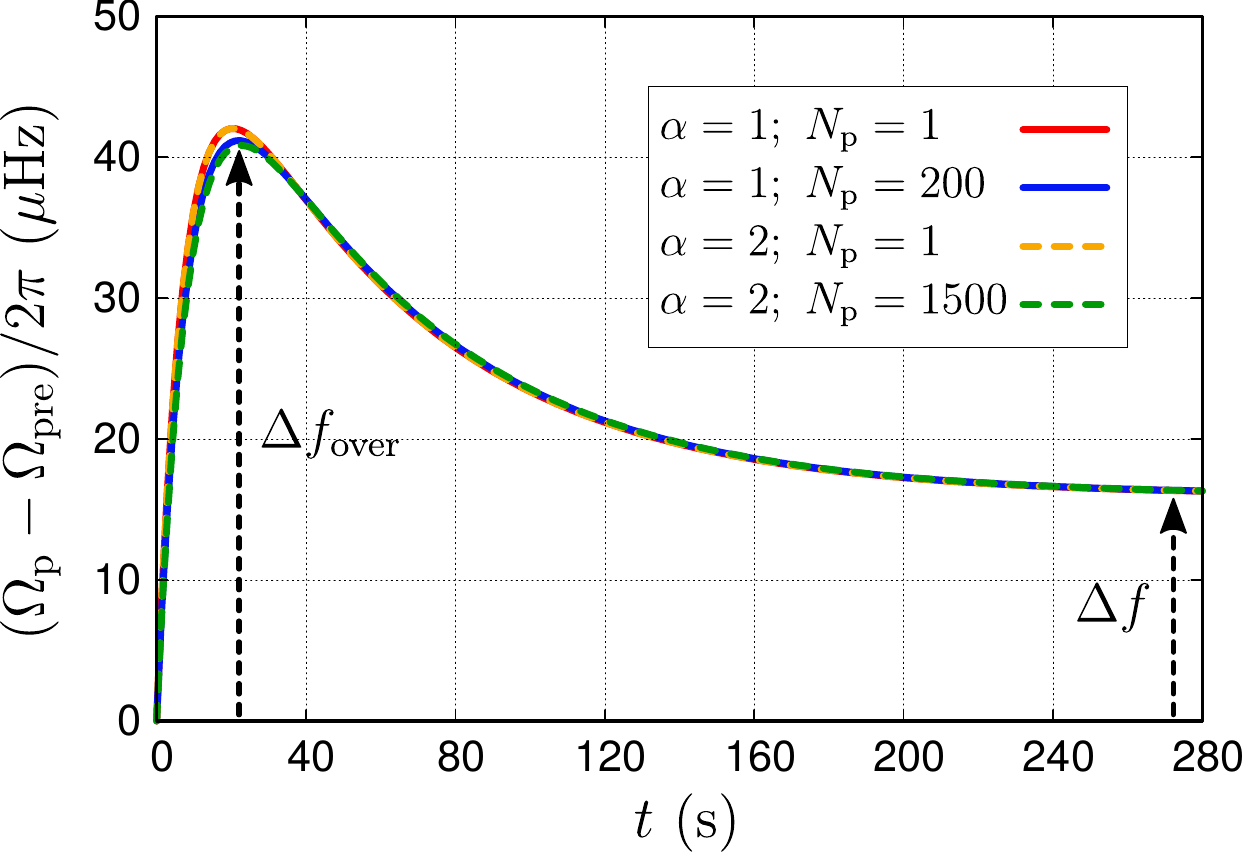}
   \hspace{0.2 cm}
   \includegraphics[width = 0.99\columnwidth]{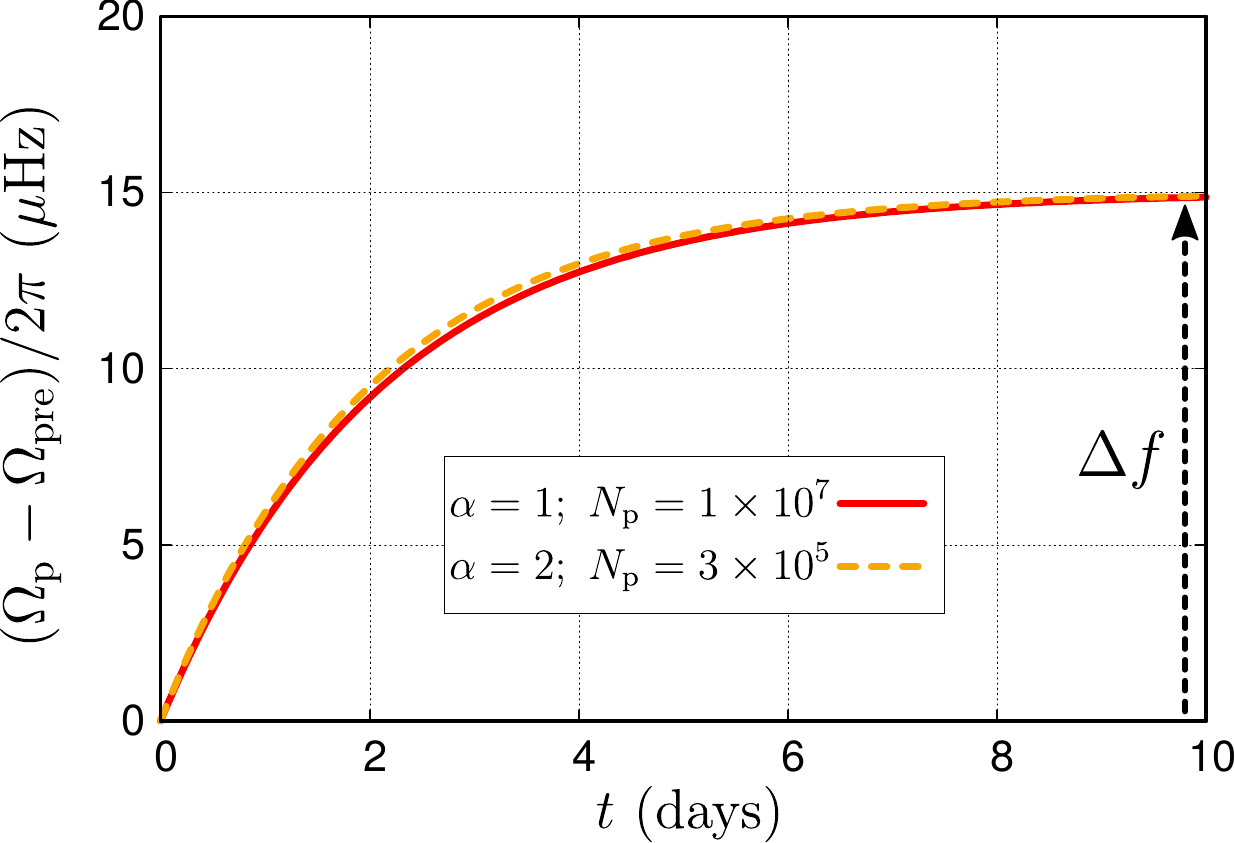} 
   \caption{(Colors online) \textbf{Left panel:} Evolution of the pulsar rotation frequency for parameters corresponding to the 2016 Vela glitch, plotted with respect to the rotation rate $\Omega_{\text{pre}}$ extrapolated from the pre-glitch evolution (i.e., in the absence of a glitch). Only the rise stage is considered. Solid (resp. dashed) lines correspond to results obtained for $\alpha=1$ (resp. $\alpha =2$). The evolution for $N_{\p}=1$ is independent of $\alpha$, see Eq.~\eqref{eq:xi_ancrage}. \textbf{Right panel:} Similar to left panel, but for input parameters corresponding to the 2017 Crab glitch. The smooth spin up and the absence of overshoot can be explained by much larger values of $N_{\p}$. See text for details. }
 \label{fig:glitch_vela_Crab}
\end{figure*}

Considering first the 2016 Vela glitch, the initial lag is fixed to $\delta \Omega_0\simeq 1.351\times 10^{-3}$~rad s$^{-1}$ so that the final glitch amplitude is $\Delta f =16$~$\mu$Hz (see Eq.~(A.16) of the SM). The evolution of the pulsar rotation frequency $\Delta\Omega_{\p}/(2\pi)$  is plotted in the left panel of Fig.~\ref{fig:glitch_vela_Crab} for $\alpha=1$, with $N_{\p} = 1$ and 200, and $\alpha=2$, with $N_{\p} = 1$ and 1500. These values lead to an overshoot of magnitude $\Delta f_\text{over}\simeq 41~\mu$Hz, a rise timescale $\tau_{\text{r}}\simeq 8$~s and a decrease timescale $\tau_{\text{d}}\simeq 57$~s, in close agreement with observations\footnote{The timescales $\tau_{\text{r}}$ and $\tau_{\text{d}}$ given for Vela correspond respectively to the quantities $\tau_-$ and $\tau_+$ introduced in the SM. For the Crab (see below), $\tau_{\text{r}}$ stands for $\tau_+$, the lower timescale $\tau_-\simeq 16.5$~s being completely negligible in this case.} (see Sec.~\ref{sec:intro}). The reason for which different values of $N_{\p}$ (for a fixed $\alpha$) lead to a similar spin-up evolution is discussed in Sec.~A2 of the SM.  For intermediate values of $N_{\p}$, the magnitude $\Delta f_\text{over}$  of the overshoot would be larger and the rise timescale $\tau_\text{r}$ would be shorter, $\tau_\text{d}$ remaining almost constant (see Figs.~A2 and A3 of the SM). Conversely, for larger $N_{\p}$, the rise time would increase and the magnitude of the overshoot would decrease until it disappears. Regarding the 2017 Crab glitch, we set $\delta\Omega_0 = 1.267\times~10^{-3}$~rad~s$^{-1}$, so that  $\Delta f= 15~\mu$Hz. The corresponding evolution of the pulsar frequency $\Delta\Omega_{\p}/(2\pi)$  is plotted in the right panel of Fig.~\ref{fig:glitch_vela_Crab} for $\alpha=1$ with $N_{\p} = 1\times 10^7$, and $\alpha=2$ with $N_{\p} = 3\times 10^5$. Such large values of $N_{\p}$ lead to a much smoother increase in the pulsar rotation rate during the glitch rise (i.e., no overshoot), with a characteristic timescale $\tau_{\text{r}}\simeq 2$ d, as observed (see Sec.~\ref{sec:intro}). 

As shown in the SM, observations of glitch overshoots set a lower bound on the moment of inertia of the non-pinned superfluid 
\begin{equation}\label{eq:Iconp-bound}
\frac{I_{\n}^{\conp}}{I}\geq 1- \frac{ \Delta f}{ \Delta f_\text{over}}\, .
 \end{equation}
The most stringent constraint so far comes from the 2004 Vela glitch~\citep{dodson2007two}, from which we deduce\footnote{We interpret the shortest timescale reported by \cite{dodson2007two} as $\tau_{\text{d}}$. In their notations, we thus have $\Delta f=\Delta F_{\p}$ and $\Delta f_\text{over} = \Delta F_{\p} + \Delta F_1$. } $\Delta f \simeq 23$~$\mu$Hz and $\Delta f_\text{over} \simeq 77$~$\mu$Hz, leading to $I_{\n}^{\conp}/I \gtrsim 0.70$. Moreover, there exists a critical value $N_{\p}^{\text{crit}, \alpha}$ of $N_{\p}$ above which no overshoot can ever occur. The presence (absence) of an overshoot in Vela (Crab) glitches thus puts constraints on the maximum (minimum) number of pinned fluxoids. 

\section{Discussions and conclusions}
\label{conclusion}

The standard scenario according to which the neutron superfluid in the core of a NS is strongly coupled to the crust on short timescales and thus cannot take part to glitch events~\citep{alpar1984rapid} must be revised if vortices are pinned to $N_{\p}$  (potentially up to $\sim 10^{13}  \, B_{12} \, P_{10}$) proton fluxoids. Using a three-component model, in which a pinned and a non-pinned core superfluids are dynamically coupled to the rest of the star through mutual friction, we have shown that the evolution of the pulsar rotation rate during the rise of a glitch can be very different depending on $N_{\p}$. While a fast spin-up with an overshoot is expected for $1\leq N_{\p}\leq N_{\p}^{\text{crit}, \alpha}$, higher values lead to a smooth rise (on a longer timescale). The value of $N_{\p}^{\text{crit}, \alpha}$ is determined by the mutual-friction coefficients. Vortex pinning can therefore account for the very different glitching behaviours observed in the Vela and Crab pulsars although the physical reason for different $N_{\p}$ remains to be investigated. The difference may lie in the spatial arrangements of fluxoids, which in turn reflect different evolutions of the internal magnetic field in these stars. More information on the internal physics of NSs can be inferred from the details of the glitch rise. In particular, observations of an overshoot set a lower bound~\eqref{eq:Iconp-bound} on the moment of inertia of the non-pinned superfluid. Allowing $N_{\p}$ to evolve may explain other observed features such as a spin-down precursor or a delayed spin-up. Vortex pinning in the outer core of NSs may thus play a crucial role, not only for the postglitch relaxation~\citep{gugercinoglu2014} but for all stages of the glitch dynamics. 

As most previous studies, our analysis was carried out in the Newtonian framework. Although general-relativistic effects may play a non-negligible role on the glitch rise~\citep{sourie2017global}, their impact remains much smaller than that of vortex pinning. Still, our treatment remains very simplified. More realistic models require a better understanding of the local dynamics of individual vortices and fluxoids, as studied, e.g., by  \cite{drummond2018}. 

\section*{Acknowledgements}

This work was supported by the Fonds de la Recherche Scientifique (Belgium) under grants no. 1.B.410.18F, CDR J.0115.18, and PDR T.004320. 



\bibliographystyle{mnras}
\bibliography{biblio}

\bsp	
\label{lastpage}
\end{document}


 
\appendix

\section{Dynamics of a glitch rise}

\subsection{Analytic solution}

Introducing the shorthand notations $i_1 = I_{\n}^{\conp} / I$, $i_2 = I_{\n}^{\cop} / I$ and $i_{\p} = I_{\p} / I = 1-i_1-i_2$, the set of equations~(9)$-$(11) of the manuscript can be simply rewritten as
\begin{align}
  & \dot{\Omega}_{\p} = -\dfrac{1}{i_{\p}}\left(i_1\, \dot{\Omega}_{\n}^{\conp}+ i_2\,\dot{\Omega}_{\n}^{\cop} + |\dot{\Omega}_\infty|\right)\, , \\
  & \dot{\Omega}_{\n}^{\conp}   = -\dfrac{i_{\p}}{\tau_1}  \left(\Omega_{\n}^{\conp} - \Omega_{\p}\right)\, ,\\
   &  \dot{\Omega}_{\n}^{\cop}   = -\dfrac{i_{\p}} {\tau_2}\left(\Omega_{\n}^{\cop} - \Omega_{\p}\right)\, , 
 \end{align}
where $|\dot{\Omega}_\infty|= -\dot{\Omega}_\infty$ and the timescales $\tau_1$ and $\tau_2$ are given by 
\begin{equation}
\label{eq:tau1_tau2}
    \tau_1 = \dfrac{i_{\p}}{2\, \mathcal{B}_{\conp} \, \Omega_{\p}}\ \ \text{and} \ \ \tau_2 = \dfrac{i_{\p}}{2\, \mathcal{B}_{\cop} \, \Omega_{\p}} \, .
\end{equation}
Note that we have neglected the very small deviations in the angular velocities in the expressions for $\tau_1$ and $\tau_2$. Unlike $\mathcal{B}_{\conp}$ (and thus $\tau_1$), the mutual-friction coefficient $\mathcal{B}_{\cop}$ in the pinned region (and therefore the coupling timescale $\tau_2$) depends on the value of $N_{\p}$ under consideration. 

Supplemented by the initial conditions provided in Sec.~3.2 of the manuscript, these equations can be solved analytically. Assuming $\tau_1$ and $\tau_2$ to be constant (which is well-justified given the  very small observed glitch amplitudes)\footnote{We recall here that $N_{\p}$ is kept fixed during the glitch rise (see Sec. 3.1 of the manuscript).}, the evolution of the pulsar rotation rate during the glitch rise is expressible as~\citep{pizzochero19} 
  \begin{equation}
      \Omega_{\p}(t) =  \Omega_\text{pre}(t) + \Delta \Omega_{\p}(t)\, ,
  \end{equation}
where $ \Omega_\text{pre}(t) = \Omega_0 + \dot{\Omega}_\infty\, t$ corresponds to the value of the  stellar angular velocity extrapolated  from the pre-glitch evolution, while the glitch deviation $\Delta \Omega_{\p}(t)$ reads
\begin{equation}
\label{eq:glitch_deviation}
\Delta \Omega_{\p}(t) = i_2\,  \delta\Omega_0 - \epsilon\, \dfrac{Q+R}{\tau_+}e^{-t/\tau_-} +  \epsilon \, \dfrac{Q-R}{\tau_-}e^{-t/\tau_+}\, ,
\end{equation}
with    
\begin{align}
Q &=1-i_1 - \beta \left(1-2\, i_1-i_2\right)\, ,   \\[0.15 cm]
R & = \sqrt{\left[1-i_1- \beta\left(1-i_2\right)\right]^2 + 4\beta i_1 i_2}\, , \\[0.15 cm]
\epsilon &= \dfrac{i_2\, \tau_1}{2\left(1-i_1-i_2\right)R} \, \delta\Omega_0\, ,\\[0.15 cm]
\tau_\pm &= \dfrac{2\, \beta\, \tau_1}{1-i_1+ \beta\left(1-i_2\right)\mp R}\, , \label{eq:tau_pm} \\[0.15 cm]
\beta &= \dfrac{\tau_2}{\tau_1} = \dfrac{\mathcal{B}_{\conp}}{\mathcal{B}_{\cop}}   \, .
\end{align}
Note that $ \tau_+ > \tau_- >0$, by definition.

\subsection{Characteristic timescales}
\label{sec:characteristic_timescales}

\begin{figure}
 \center 
   \includegraphics[width = \columnwidth]{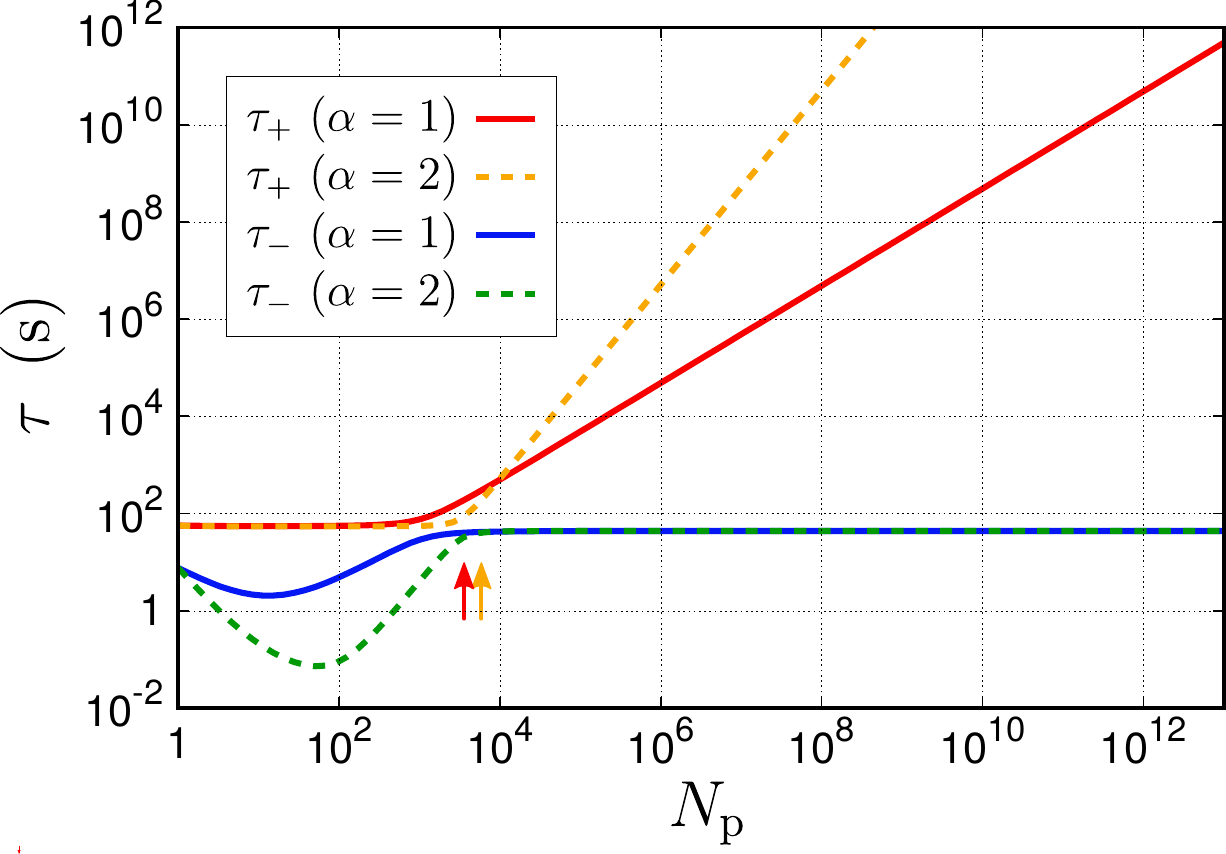}
   \caption{Characteristic timescales $\tau_+$ and $\tau_-$~\eqref{eq:tau_pm} as functions of $N_\text{p}$ for $\alpha = 1$ (solid lines) and $\alpha =2$ (dashed lines). The red (resp. orange) vertical arrow indicates the value $N_{\p}^{\text{crit}, 1}$ (resp. $N_{\p}^{\text{crit}, 2}$) of $N_{\p}$ for $\alpha = 1$ (resp. $\alpha= 2$). See text for details. }
 \label{fig:characteristic_timescales}
\end{figure}

The characteristic timescales $\tau_+$ and $\tau_-$~\eqref{eq:tau_pm} governing the evolution of the pulsar rotation rate during the glitch rise~\eqref{eq:glitch_deviation} are plotted in Fig.~\ref{fig:characteristic_timescales} for the two prescriptions $\alpha = 1$ and $\alpha = 2$ considered in the manuscript (see Sec.~2.2). The rotation rate $\Omega_0$ is set to that of the Vela pulsar, i.e., $\Omega_0 = 2\pi \times 11.19$~rad~s$^{-1}$. The following values are taken for the proton fraction, total density and proton entrainment parameter in the pinned region:  $x_{\p}^{\cop} = 0.07$, $\rho_{\cop} = \rho_0 =  2.7\times 10^{14}~\text{g~cm}^{-3}$ and $\varepsilon_{\p}^{\cop}=0.05$. The same quantities in the non-pinned region are fixed to: $x_{\p}^{\conp} = 0.2$, $\rho_{\conp} = 3 \rho_0$ and $\varepsilon_{\p}^{\conp}=0.1$. The ratios $I^{\cru}/I$ and $I^{\cop}/I$ are respectively set to $0.03$
 and $0.08$.  For each $\alpha$, the critical value $N_{\p}^{\text{crit}, \alpha}$ of $N_{\p}$ for which $\beta= 1$ is indicated by an arrow in Fig.~\ref{fig:characteristic_timescales}. Values of $N_{\p}$ between $1$ and $N_{\p}^{\text{crit}, \alpha}$ correspond to $\beta\leq 1$ (see also Sec.~\ref{sec:overshoot}). 
 
 As can be seen in Fig.~\ref{fig:characteristic_timescales}, the upper timescale $\tau_+$ is nearly constant ($\tau_+\lesssim 100$~s) for $1\leq N_{\p}\leq N_{\p}^{\text{crit}, \alpha} $, whereas it strongly increases with  $N_{\p}$ for $ N_{\p} > N_{\p}^{\text{crit}, \alpha} $, reaching values possibly longer than months or years. Conversely, the lower timescale $\tau_-$ is found to be highly dependent on $N_{\p}$ for $1\leq N_{\p}\leq N_{\p}^{\text{crit}, \alpha} $, but is almost constant  for higher $N_{\p}$. In particular, there always exists two different values of $N_{\p}\leq N_{\p}^{\text{crit}, \alpha}$ leading to the same $\tau_-\leq \tau_-\left(N_{\p}=1\right)$. The origin of this degeneracy is to be found in the profile of $\mathcal{B}_{\cop}$ at small $N_{\p}$ (see Fig.~1 of the manuscript).

 The previous dependencies can be understood by looking at the two limiting cases $\beta \ll 1$ and $\beta \gg 1$. For $\beta \ll 1$, the characteristic timescales $\tau_+$ and $\tau_-$ reduce to 
\begin{equation}
\tau_+ \simeq \tau_1 \dfrac{ 1-i_1}{i_{\p}} =\dfrac{1-i_1}{2\, \mathcal{B}_{\conp} \, \Omega_{\p}}\, , 
\end{equation}
and
\begin{equation}
\tau_- \simeq \tau_2   \dfrac{1}{1-i_1} = \dfrac{i_{\p}}{\left(1-i_1\right)2\, \mathcal{B}_{\cop} \, \Omega_{\p}} \, .
\end{equation}
In the opposite limit $\beta\gg 1$, $\tau_+$ and $\tau_-$ simplify as 
\begin{equation}
    \tau_+ \simeq \tau_2  \dfrac{1-i_2}{i_{\p}} = \dfrac{1-i_2}{2\, \mathcal{B}_{\cop} \, \Omega_{\p}} \, ,
\end{equation}
and 
\begin{equation}
    \tau_- \simeq \tau_1  \dfrac{1}{1-i_2}=  \dfrac{i_{\p}}{\left(1-i_2\right)2\, \mathcal{B}_{\conp} \, \Omega_{\p}}\, .
\end{equation}
Since $\mathcal{B}_{\conp}$ does not depend on $N_{\p}$ (unlike $\mathcal{B}_{\cop}$), $\tau_+$ (resp. $ \tau_-$) is found to be independent of $N_{\p}$ in the limit $\beta \ll 1$ (resp. $\beta \gg 1$), i.e., for values of $N_{\p}$ around the minimum of $\tau_-$ (resp. for $N_{\p}\gg N_{\p}^{\text{crit}, \alpha} $), as observed in Fig.~\ref{fig:characteristic_timescales}. In the absence of pinning, $\tau_+ \simeq 3426$~s and $\tau_-\simeq 43.5$~s.

\subsection{Overshoot: condition and upper limit}
\label{sec:overshoot}

From Eq.~\eqref{eq:glitch_deviation}, we deduce that the jump $\Delta \Omega$ in the pulsar rotation rate at the end of the glitch ($t\rightarrow + \infty$) is given by
\begin{equation}
    \Delta \Omega = i_2 \, \delta\Omega_0 \, .
    \label{eq:final_amplitude}
\end{equation}
As discussed in~\cite{pizzochero19}, the pulsar rotation rate during the glitch rise can actually exceed the final glitch amplitude $\Delta\Omega$ if some specific condition is fulfilled. Such an overshoot is found to occur at the time \begin{equation}
t_\text{over} = \dfrac{\tau_1 \beta}{R} \ln{\left(\dfrac{Q+R}{Q-R}\right)}\, , 
\label{eq:t_over}
\end{equation}
provided that $Q>R$ or, equivalently, $\beta<1$. The condition $\beta=1$ determines a critical value $N_{\p}^{\text{crit}, \alpha}$ of $N_{\p}$ above which no overshoot can ever occur. The adopted parameters yield  $N_{\p}^{\text{crit}, 1}\simeq 3589$ and $N_{\p}^{\text{crit}, 2}\simeq  5869$. Assuming $X\gg 1$ and $\xi_0^{\conp}\ll 1$ with $\gamma\equiv \left(\varepsilon_{\p}^{\cop}\right)^2 \,  \xi^{\cop}_0/\xi^{\conp}_0$ and $y_{\p}\equiv \left(\varepsilon_{\p}^{\cop}\, x_{\p}^{\cop}\right)^2/\left(1- x_{\p}^{\cop}\right)^2$, we find 
\begin{equation}
    N_{\p}^{\text{crit}, 1}=\dfrac{\gamma}{\left(\xi^{\cop}_0\right)^2+y_{\p}}\, , \hskip 0.5cm N_{\p}^{\text{crit}, 2}=\dfrac{\sqrt{\gamma-y_{\p}}}{\xi^{\cop}_0}\, ,
\end{equation}
$\xi^{\cop}_0$ and $\xi^{\conp}_0$ being the drag-to-lift ratios for $N_{\p}=0$ in the pinned and non-pinned regions, respectively (see Eq.~(8) of the manuscript). These expressions reproduce the numerical results with an error of about $0.7\%$ ($6 \times 10^{-5}\%$) for $\alpha=1$ ($\alpha =2$). No overshoot occurs for $N_{\p}= 0$ since the condition $\beta\approx \xi^{\conp}_0/\xi^{\cop}_0>1$ is satisfied for the adopted parameters. 

\begin{figure}
 \center 
   \includegraphics[width = \columnwidth]{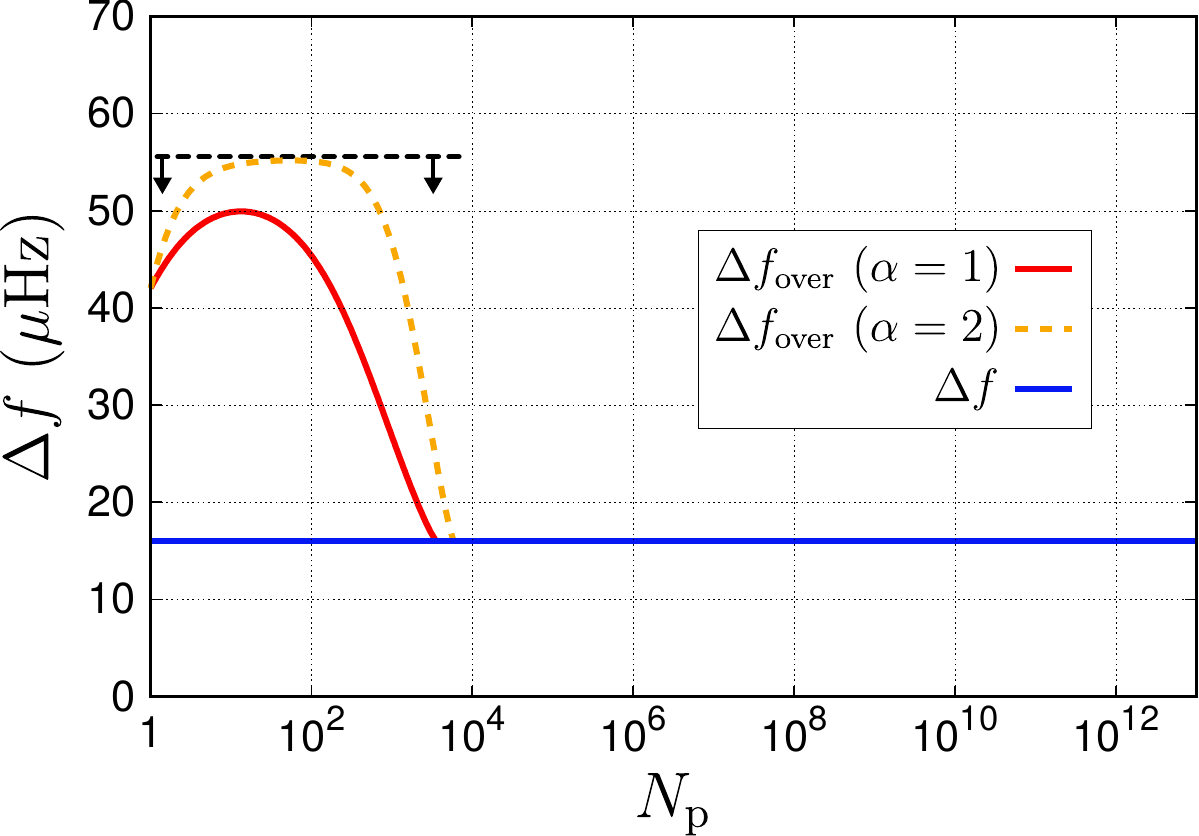} 
   \caption{Amplitude $\Delta f_{\text{over}}$ of the overshoot as a function of $N_{\p}$ for $\alpha = 1$ (red solid line) and $\alpha=2$ (orange dashed line). The final glitch amplitude $\Delta f = \Delta \Omega / (2\pi)$~\eqref{eq:final_amplitude} is indicated by an horizontal blue line and corresponds to that observed in the 2016 Vela glitch~\citep{asthon19rotational}. The upper limit~\eqref{eq:upper_ampl_over} is also displayed with an horizontal dashed line. See text for details.}
 \label{fig:overshoot}
\end{figure}

The amplitude  $\Delta f_\text{over}\equiv \Delta \Omega_{\p}(t_\text{over})/(2\pi)$ of the overshoot, see Eqs.~\eqref{eq:glitch_deviation} and~\eqref{eq:t_over}, is plotted with respect to $N_{\p}$ in Fig.~\ref{fig:overshoot}. The initial lag is fixed to $\delta \Omega_0\simeq 1.351\times 10^{-3}$~rad s$^{-1}$ so that the final glitch amplitude~\eqref{eq:final_amplitude} is $\Delta f = \Delta \Omega / (2\pi)=16$~$\mu$Hz, as observed in the 2016 Vela glitch~\citep{asthon19rotational}. As shown above, an overshoot only exists if $1\leq N_{\p}< N_{\p}^{\text{crit}, \alpha}$. Since the amplitude $\Delta f_\text{over}$ of the overshoot is found to increase with decreasing $\beta$, its maximum possible value is thus determined by the limit $\beta \ll1$: 
\begin{equation}
     \Delta f_\text{over} \leq  \dfrac{\Delta f}{1-i_1}\, , 
     \label{eq:upper_ampl_over}
 \end{equation}
as shown in Fig.~\ref{fig:overshoot}. Using the normalization of the  moments of inertia, the inequality~\eqref{eq:upper_ampl_over} also leads to
 \begin{equation}
    i_2 \geq \left(1-\dfrac{I^{\cru}}{I}- \dfrac{i_1}{1-x_{\p}^{\conp}}\right)\left(1-x_{\p}^{\cop}\right) \, .
\end{equation}
Values of the model parameters can be extracted by fitting the full evolution of the pulsar rotation rate~\citep{pizzochero19}.

\begin{figure}
 \center  
 \includegraphics[width = \columnwidth]{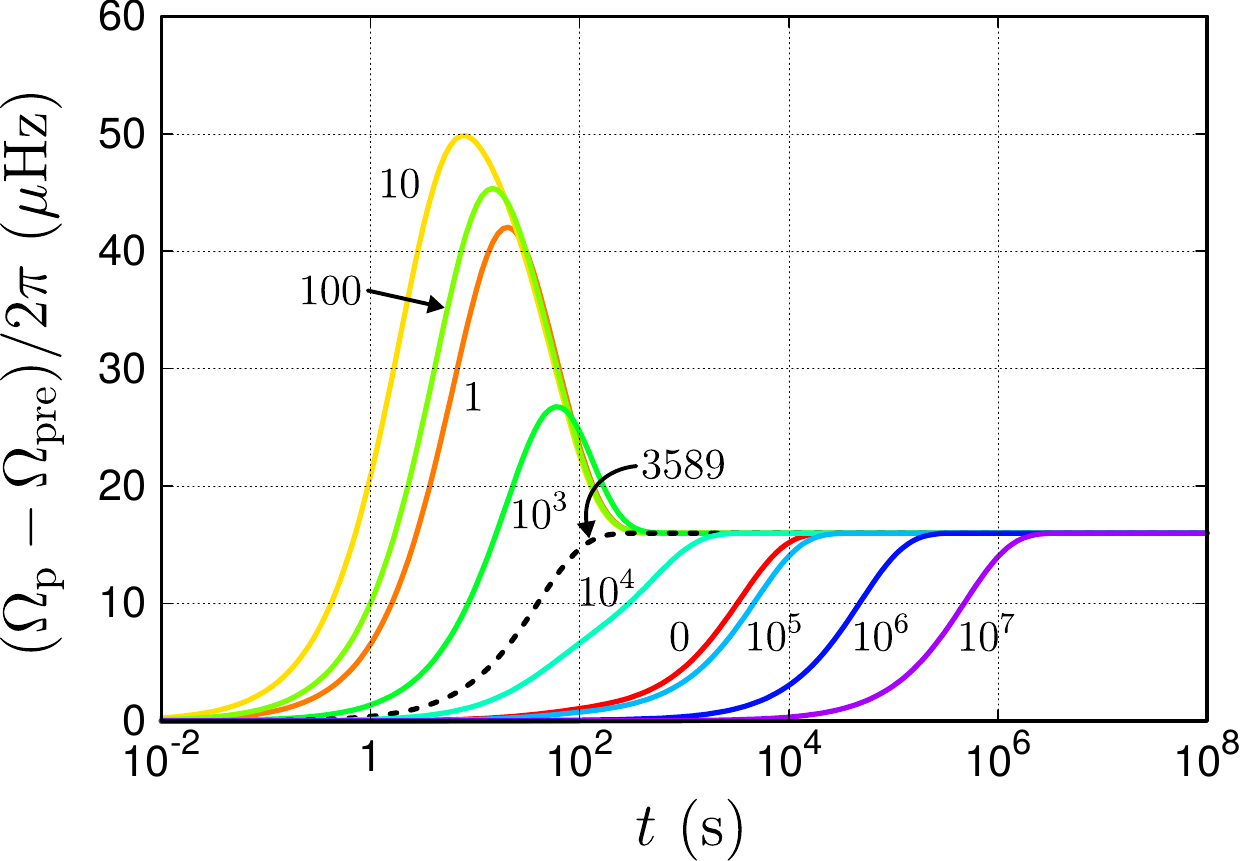} 
   \caption{Evolution of the pulsar frequency $\Delta \Omega_{\p} /(2\pi)$, Eq.~\eqref{eq:glitch_deviation}, during the glitch rise for different values of $N_{\p}$ using $\alpha = 1$. Numbers on the curves represent the values for $N_{\p}$. }
 \label{fig:Np}
\end{figure}

The evolution of the pulsar frequency~$\Delta \Omega_{\p}/(2\pi)$ during the glitch rise, as given by Eq.~\eqref{eq:glitch_deviation}, is plotted in Fig.~\ref{fig:Np} for different values of $N_{\p}$. Although only $\alpha=1$ is considered, results for $\alpha=2$ are qualitatively similar. As previously discussed, an overshoot only occurs if $1\leq N_{\p}< 3589$, while a much smoother increase in the rotation rate is observed for values of $N_{\p}$ outside this range, with a characteristic rise timescale corresponding to $\tau_+$.

\bibliographystyle{mnras}
\bibliography{biblio}

\bsp	
\label{lastpage}